\documentclass[10pt,journal,compsoc]{IEEEtran}
\ifCLASSOPTIONcompsoc
  \usepackage[nocompress]{cite}
\else
  \usepackage{cite}
\fi
\usepackage{epigraph}
\usepackage[pdftex]{graphicx}
\ifCLASSINFOpdf
\else
\fi
\usepackage{algorithm, algorithmic}
\usepackage{hyperref}

\usepackage[switch]{lineno} 

\begin{document}
%
\title{An Agent-Based Modelling Approach to Brain Drain}
%
%
%
%

\author{Furkan~Gürsoy~\IEEEmembership{}    and~Bertan~Badur~\IEEEmembership{}
\IEEEcompsocitemizethanks{
\IEEEcompsocthanksitem F. Gürsoy and B. Badur are with the Department of Management Information Systems, Boğaziçi University, Bebek, Istanbul, 34342, Turkey. (e-mail: furkan.gursoy@boun.edu.tr)}
\thanks{ \url{https://doi.org/10.1109/TCSS.2021.3066074} \copyright 2021 IEEE.}
}

\IEEEtitleabstractindextext{%
\begin{abstract}
The phenomenon of brain drain, that is the emigration of highly skilled people, has many undesirable effects, particularly for developing countries. In this study, an agent-based model is developed to understand the dynamics of such emigration. We hypothesise that skilled people's emigration decisions are based on several factors including the overall economic and social difference between the home and host countries, people's ability and capacity to obtain good jobs and start a life abroad, and the barriers of moving abroad. Furthermore, the social network of individuals also plays a significant role. The model is validated using qualitative and quantitative pattern matching with real-world observations. Sensitivity and uncertainty analyses are performed in addition to several scenario analyses. Linear and random forest response surface models are created to provide quick predictions on the number of emigrants as well as to understand the effect sizes of individual parameters. Overall, the study provides an abstract model where brain drain dynamics can be explored. Findings from the simulation outputs show that future socioeconomic state of the country is more important than the current state, lack of barriers results in a large number of emigrants, and network effects ensue compounding effects on emigration. Upon further development and customisation, future versions can assist in the decision-making of social policymakers regarding brain drain.
\end{abstract}

}

\maketitle

\IEEEdisplaynontitleabstractindextext

%
\IEEEpeerreviewmaketitle

\IEEEraisesectionheading{\section{Introduction}\label{sec:introduction}}

\IEEEPARstart{B}{rain} drain can be defined as the emigration of highly skilled or qualified people from a particular country.  It has many undesirable effects for the home (i.e., source, sender) countries since the loss of human capital damages the economic and social development \cite{drainorbank, Berry1969}. On the other hand, host (i.e., target, receiver) countries often benefit from it \cite{chojnicki2004economic, jaumotte2016migrants}. For developing countries, the flow of their skilled labour to developed countries is a major concern, or at least it should be. Purpose of this study is to explore the dynamics of migration of human capital which is an important issue, especially for developing countries. Ideally, such countries would want to develop policies to keep the skilled workforce in the country so that a sustainable national economy can be achieved.

An analytical solution for the defined problem would be difficult, if not impossible, to develop and also difficult to understand. Since our problem includes coordination between the agents and complex behaviours, an agent-based modelling (ABM) \cite{wilensky2015introduction,railsback2019agent} approach is more useful. Moreover, employing an ABM approach provides better flexibility in terms of modifying the model structure and agent behaviour \cite{hammond2015considerations}. For these reasons, an agent-based model is developed to understand the dynamics of brain drain. The model we propose can be further improved to serve as a decision support tool and visual demonstration tool for policymakers. 

In this study, we focus on emigration of highly skilled workers rather than overall migration. Migration might happen for many reasons such as climate change, natural disasters, war, population exchange treaties, political dissidence and oppression, seeking higher education, government programs/invitation for unskilled laborers from prespecified countries, former colonial relationships, rural-urban dynamics, family reunions and so on. Brain drain has different characteristics in terms of underlying reasons, enabling factors, migration process, decision-makers, sociological and political aspects, etc. In addition, the research questions and aspects which the society is concerned are often different for different types of migrations. Some are more relevant for studying the impact of climate change, others are for integration issues, international relations, prevention of mass migration, etc.  In summary, different types of migration require different modeling decisions even in the relatively abstract models because the underlying factors, processes, and outcomes of interest differ. 

Based on the relevant literature outlined in the next section, we hypothesise that there are two main factors which impact the decision and ability to move abroad: desire to maximise individual utility and network effects. Accordingly, several factors play role in brain drain such as the overall economic and social differences between the home and host countries, people's ability and capacity to obtain good jobs and start a life abroad, the barriers of moving abroad, and people's social network who are already working abroad. This is also in line with the four factors playing role in migration in a pull-push framework: factors associated with the home, factors associated with the host, intervening obstacles, and personal factors \cite{leetheory}.

One's utility in moving abroad might be economical, social, or a combination of the two. Throughout the study, the term \textit{utility} is used in reference to change in utility in case of moving abroad. Change in economic utility can be measured by comparing income prospects, cost of living, and other relevant information. On the other hand, social utility is more difficult to measure due to its qualitative nature. Family relations, the political climate in the country, individual political orientations, lifestyle choices, and so on all play a role in emigration wish. In this study, the total individual utility is measured in a more abstract level by using a single variable. Such abstraction is preferred since modelling utility in a more physical level with sub-utilities requires very extensive qualitative and quantitative data collection and analysis effort, which should ideally be a long-continued collaborative effort of a number of researchers with diverse specialisations.

Network effects play an important role in facilitating one's emigration by increasing the chances of finding employment abroad, providing social support, etc. But how such network effects can be calculated? The first approach is that the whole population from the same nation living in the target country can be considered as an aggregate variable. Although this might be a good approximation for some problems, it does not help to distinguish the individuals which have different network characteristics. The second approach proposes that the social network effects are calculated based on ego networks of individuals which only include meaningful connections. We believe that the latter provides a better approximation of real-world mechanisms; therefore, a social network approach is employed in this work.

The rest of this study is organised as follows. The relevant literature on agent-based approach to brain drain and migration in general is briefly visited in the next section. In Section 3, the proposed agent-based model is described in detail. In Section 4, qualitative and quantitative real-world patterns regarding brain drain, which we build and calibrate the model upon, are presented. Using the calibrated model as the reference; results of extreme parameter values and existence of tipping points are studied, local and global sensitivity analyses and uncertainty analyses are performed, robustness to different network types is investigated, and a response surface analyses with linear and random forest models are presented in Section 5. Final remarks and conclusions are given in Section 6 along with directions for future research.

\section{Related Work}

The literature on agent-based models of emigration of skilled workers (i.e., brain drain) is very scarce. Gorrieri \cite{gorrieri} develops an agent-based simulation model to understand the functioning, decisional factors, and consequences of brain drain. Specifically, it consists of two regions (i.e., home country and elsewhere abroad) and three components: agents' decision to study in university, job search after graduation and possible emigration, and return migration. At the beginning of the simulation, all agents are born at age 19 and decide to study according to an increasing threshold function of their ability, number of friends in higher education, and the relative economic performance of their country. If they spend a semester studying abroad, they form a new set of friendships there. Once they graduate, they start searching for jobs and move abroad based on random personal traits, diploma convertibility, and available jobs which is an increasing function of economic growth. Among those who emigrated, they might eventually return depending on the number of their friends abroad, personal traits, and job availability at home. They present the effects of network density, diploma convertibility, and economic differences on the simulation results.

Biondo et al. \cite{biondo2013} study the return migration after brain drain. They assume that agents are already emigrated and approach the problem from a rational microeconomic perspective and by modelling the individuals' rational decisions to return after emigrating to another country. The network effects are included via simple star-like networks at home and host countries. The decision is based on individuals' risk aversion (i.e., representing personal traits) and initial expectation (i.e., representing the anticipated utility abroad). Their main finding is that the return migration strongly depends on the ratio of risk aversion to initial expectation. The authors also mention the lack of migration data for skilled people, which is necessary to have a non-abstract model that can be operationalized.

There are also studies that focus on the emigration in specific high-skill occupations. Vaccario et al. \cite{vaccario2018reproducing} build an agent-based model of global scientist movements where migration decisions are based on geographical distance and prestige of academic institutions.  Amorim-Lopes et al. \cite{Amorim-Lopes2019} build an agent-based model for lifecycle and emigration of physicians in Portugal. Both studies are examples of more physical rather than abstract models and very specific rather than general brain-drain models. In both, the agents emigrate to maximize a utility, operationalized as prestige for the scientists (based on the assigned scores for publications) and by public and private goods for the physicians. Geographical distance and migration cost play adverse roles against emigration respectively in two studies. On the other hand, network effects are missing in both studies.

The ABM literature on migration in general is more developed. Silveira et al. \cite{Silveira2006445} develop a model with a statistical physics approach inspired by the Ising model to study the migration between urban and rural areas in the early phases of industrialisation. The migration decision is mainly based on job availability, wage differences, and influence from the social neighbourhood. Fu and Hao \cite{fu2018agent} find that incorporating social networks improves the performance of migratory decision models and such networks serve as catalysts in guiding rural-urban migration. Barbosa Filho et al. \cite{Filho2011109} study the influence of social networks on migration by employing an evolutionary perspective. Kniveton et al. \cite{Kniveton2011, Kniveton2012444} employ the theory of planned behaviour (TPB) \cite{ajzen1991} for migration decision-making and explore the role of climate change in migration in Burkina Faso, in addition to the population growth rate and socioeconomic factors. Klabunde et al. \cite{Klabunde201751} develop an agent-based model inspired by TPB for migration from Senegal to France, also considering social network links. Hassani-Mahmooei and Parris \cite{hassani2012climate} study internal migration patterns in Bangladesh and forecasts migration dynamics as a result of climatic and socioeconomic factors. Similarly, Smith \cite{Smith201477} presents an agent-based model to study and forecast migration in Tanzania where people migrate based on environmental factors such as rainfall and also demographic and societal changes.

Although a simulation model is not developed, Haug \cite{Haug2008585} discusses the role of location-specific social capital in migration decision-making based on survey data from Germany and Bulgaria. Its findings indicate that social capital at destinations increases the emigration or return migration to those destinations. Similarly, Kley \cite{Kley2011469} analyzes the longitudinal surveys of German participants to explore the migration decision-making and behaviour and finds that perceived opportunity differentials and influence of significant others are among important factors. 

Generally, apart from the personal traits, the factors affecting the migration in general can be grouped under two broad categories: economic utility (including the physiological needs) and the network effects. The first one is supported by the traditional economic theories such as rational choice \cite{scott2000rational} and relative utilitarianism \cite{dhillon1999relative}. The network effects, on the other hand, are supported by different theories that usually argue for the effects of affinity, information, and facilitation as well as by empirical evidence \cite{Haug2008585, ORRENIUS2005215}. Often, the network effects are considered broadly as social capital without explicitly modelling the underlying network structure. Overall, it can be concluded that a migration model should account for economic utility and network effects.

From an empirical point of view, the lack of real-world data restricts the development of less abstract agent-based models of brain drain. Although public datasets are available on migrant numbers between countries \cite{wbdata, undata, oecddata}, no information is available on the educational or skill level breakdown of these migrants. Biondo et al. \cite{biondo2013} also report that availability of such data would enable better model parameter calibration and they are working to construct such database. However, to the best of our knowledge, real-world data on migration numbers of highly skilled workers is not publicly available as of now. A very recent International Organization for Migration report \cite{iomdata} also states that their statistics do not reflect policy categories such as highly-skilled workers. For the most part, this prohibits developing more physical models which can utilize available real-world information (e.g., macroeconomic variables, social indices, personal surveys, etc.) that can quantitatively explain the real-world dynamics of brain drain. Nevertheless, the main benefit of a policy model is the insights into the policy domain rather than the numbers it generates and modelling is still valuable even if data calibration is not possible \cite{gilbert2018}. Even more, Epstein \cite{epstein2008} enumerates 16 reasons other than prediction to build models.

Given the limited literature on agent-based models of brain drain and lack of relevant comprehensive public data, the contribution of our paper, apart from the findings from the simulation experiments, can be summarized as follows:
\begin{itemize}
    \item We provide an abstract model specifically for brain drain which can serve as a baseline model. Different scenarios are simulated using the proposed model and findings are presented. Future research can easily replace that the abstract variables in our model with functions of a set of observed variables and test whether these variables can explain the brain drain dynamics, once relevant public datasets are available. We also mention a few extension ideas that can be considered in future research.
    
    \item We obtain real-world data, albeit a limited one, and calibrate our model parameters accordingly. This enables model analyses (e.g., sensitivity analysis, uncertainty analysis) to be done with a realistic baseline model.
    
    \item In addition to network density (i.e., the total number of links), this study employs two different network structures and compares their effects on the brain drain dynamics.
        
    \item The impact of employed variables on the brain drain are investigated and their effect sizes and statistical significance are discussed utilizing a linear response surface model. Moreover, a machine learning model that is capable of accurately predicting the simulation output is presented.
    
\end{itemize}

\section{The model}

The proposed agent-based model is described in this section, following the \textit{overview, design concepts, and details (ODD)} protocol of Grimm et al. \cite{grimm2006, grimm2010}. The detailed and structured description ensures that the model is easily understandable and reproducible. The source code is made available online\footnote{The model in Netlogo can be found in CoMSES Computational Model Library at \url{https://doi.org/10.25937/6cfk-dk31}}.

\subsection{Purpose}

Brain drain can be defined as the emigration of highly trained or qualified people from a particular country, which has many undesirable effects for the source country. Purpose of the model is to understand the dynamics of brain drain; and provide an initial version of a simulation-based decision support tool which can be used in discovering future trends for such emigration, and design effective social policies which can reduce, stop, or reverse the brain drain. The model proposes that skilled people would like to emigrate to maximise their utility, yet actual emigration is constrained with barriers, luck, and individuals' social network.

\subsection{Entities, state variables, and scales}

There exists only a single type of entity, agents, to represent individual persons who are well educated/skilled. Each person is characterised by a number of states: years spent in the workforce ($age$), whether the agent has moved abroad or still in the country ($s$), initial utility representing the agents' utility of moving abroad at the beginning of a simulation ($x$), and the utility of an agent during the course of simulation ($y$) which is calculated based on initial utility and relative attractiveness difference between host and home countries ($G$). The values for $x$ and $y$ are relative rather than absolute, with positive values indicating that the target country provides more utility than the source country. The actual probability of moving abroad consists of a component ($p$) and the luck of the agent ($luck$).  $p$ is a function of utility and the number of network-neighbours living abroad (i.e., $s = 0$).

The source and target countries are not explicitly defined as separate entities. The model assumes only a single source country and a single target country which represents abroad in general. Therefore, there implicitly exists a type of entity other than the individual persons. There are global variables which correspond to the relative or absolute states of the source country. Barriers to moving abroad ($threshold$) and relative attractiveness difference between host and home countries ($G$) are such variables. Larger values of $G$ indicate that the target country is more attractive than the sender country in general. $G_t$ is its value at $t=t$, $G_0$ is its initial value at $t=0$ and $G^*$ is its final value in the long run.

Additional global variables in the model are as follows. The number of meaningful connections on average an individual has in the network may potentially increase one's chances of moving abroad and is calculated using $linksToNodesRatio$. The network effect coefficient ($networkCoef$) controls the weight of network effects against the utility effect. Luck reduction coefficient ($luckCoef$) determines how much an individual's $luck$ is reduced in case of unsuccessful emigration attempt.

The model works at discrete time steps and each time step corresponds to a year. The model is simulated for 50 time steps since the authors believe that a period of 50 years is long enough but not too long for obtaining meaningful insights from such model as ours.

\subsection{Process overview and scheduling}

At each time step, the following tasks occur in the order of their appearance here. 

\begin{enumerate}
    \item $age$ of all agents are incremented by one. 
    \item Certain number of agents die, and the same number of agents are born.
    \item $G_t$, that is the attractiveness difference between the source and target countries, is updated.
    \item $y$, that is the utility of moving abroad, of all agents are updated. 
    \item $p$ values of all agents are updated to reflect the changes in $y$.
    \item Agents who satisfy the $p>threshold$ condition migrate with a probability equal to $p*luck$. 
    \item For individuals who successfully migrated, $s$ is set to $0$. For individuals who attempted but failed to migrate, $luck$ is updated.
\end{enumerate}

\subsection{Design Concepts}

\textbf{Basic Principles.} This model assumes a simple utilitarian approach with barriers and random processes for actually performing the most beneficial action: staying in the country or moving abroad. Network effects also play a role in migration by making it more likely for people to move abroad if they have more people in their network who live abroad. Moreover, unsuccessful emigration attempts reduce future chances of emigration. Each unsuccessful attempt uncovers new information indicating further barriers for that agent's emigration. If this is omitted, too many agents who have positive utility eventually emigrate in one of their large number of repetitive attempts over the course of the simulation. 

\textbf{Emergence.} Numbers of agents in the source and target countries emerge from the value of $G$ that represents the general socioeconomic status difference between the source and target countries, and $threshold$ which indicates the degree of the barriers against moving abroad. Then, essentially, there are two control mechanisms available for policy-makers: either increasing the barriers of moving abroad (which might result in undesirable side effects depending on the nature of the policies) or designing policies towards a better socioeconomic environment.

\textbf{Sensing.} Individuals have complete information on $G$, $threshold$, and the status ($s$) of their network-neighbours.

\textbf{Interaction.} An individual's chance of moving abroad increases as they have more network-neighbours, which can be considered as an implicit form of interaction.

\textbf{Stochasticity}. Actual emigration is a simple stochastic process based on the probability of emigration ($p*luck$) of each agent, constrained by the specified $threshold$ value.

\textbf{Observation.} A plot is used for observing the number of agents staying in the country over the course of the simulation. In addition, two simple monitors are used to display the number of individuals in the home country and abroad. A visual display is also included which displays the links between the agents who are represented with different colours based on whether they live in the home country or abroad.

\subsection{Initialisation}
Initially, $1000$ agents are created. All $s$ values are initiated as $1$ (i.e., all agents are born in the home country). All agents have the same initial $luck$ value of $1$. Their $age$ values are randomly assigned between $0$ and $30$, assuming that an agent stays in the workforce for 30 years. Finally, all agents have $x$ values uniformly distributed between $-1$ and $1$, which means that some have better utility in moving abroad and others in staying in the country.

\subsection{Input Data}
The model does not use input from external models or data files.

\subsection{Submodels}

Due to initialisation of 1000 agents aged uniformly in $0-30$ range, around 33 people would have $age$ value of $30$. In order to ensure the initial assumption that an agent stays in the workforce for around 30 years, 33 agents who have the greatest age are selected and removed from the model (i.e., death of agents). In case of more than enough number of agents having the same age, the agents are randomly selected among them. For the birth of agents; 33 agents are duplicated, creating 33 agents with the same state values. However, we initialise all state values and network links from scratch; therefore, there is effectively no duplication. (In the earlier versions, duplication is used to carry some attributes of the parent agents to the newborn agents. In this version, we do not require such state value preservation. However, we chose to keep it this way in the code so that future versions make modifications on it to achieve such value preservation more easily.). $luck$, $age$,  $x$ are assigned values same as the original initialisation. For each newborn agent, an appropriate number of links are created based on the selected network type. The expected value of the number of links in the network does not change over time. 

Equation \ref{eq:1} shows the formula for how $G$ gradually attains the value of $G^*$ based on its initial value at $t=0$ that is $G_0$, its value in the long run that is $G^*$, and the current time step $t$. The decay parameter $d$ is set as $-0.1$ in this study.

\begin{equation}\label{eq:1}
G_t = G^*  - (G^* - G_0) * e^{(d*t)}
\end{equation}

$y$ values are updated according to Equation \ref{eq:2}.
\begin{equation}\label{eq:2}
y_t = x + G_t
\end{equation}

$p$ values are updated based on Equation \ref{eq:3}, given that $NA$ is the set of network-neighbours who already live abroad. The equation consists of two components: utility effects and network effects. Utility values are scaled and shifted by respective parameters $scale$ and $shift$ and used in a logistic function to produce utility effects. Network effects are calculated via the natural logarithm operation. A coefficient ($\alpha$) is utilised to balance range differences between utility effects and network effects. Initial experiments during model development showed that when $scale$, $shift$, and $\alpha$ are set values of $1.5$, $4$, and $0.05$ respectively, realistic and meaningful $p$ values were obtained. Hence, those parameter values are employed in this study. 

\begin{equation}\label{eq:3}
p = \frac{e^{scale*y-shift}}{e^{scale*y-shift}+1} + \alpha*networkCoef * ln(1+|NA|)
\end{equation}

The emigration process is shown in Algorithm 1. It also demonstrates how the values of $s$ and $luck$ are updated.

\begin{algorithm}[H]
\caption{Emigration attempt}
\begin{algorithmic}
\IF{$p > threshold$}
    \IF{p * luck $>$ random(0,1)}
        \STATE $s \gets 0$
    \ELSE    
        \STATE $luck \gets luck * luckCoef$
    \ENDIF
\ELSE 
    \STATE $continue$
\ENDIF
\end{algorithmic}  
\end{algorithm}\label{alg:1}

\section{Pattern-oriented modelling and model calibration}\label{sec:pom}

\setlength{\epigraphwidth}{0.45\textwidth}
\epigraph{You can catch phenomena in a logical box or in a  mathematical box. The logical box is coarse but strong. The mathematical box is fine-grained but flimsy. The mathematical box is a beautiful way of wrapping up a problem, but it will not hold the phenomena unless they have been caught in a logical box to begin with.}{\textit{Platt \cite{platt1964}}}

Qualitative pattern matching often comes before a quantitative calibration, as argued in \cite{platt1964, grimm2012}. To this respect, a visual pattern-matching strategy is continuously employed during the model development to satisfy the following qualitative pattern statements.  Ceteris paribus, more people move abroad when

\begin{itemize}
    \item the benefits of working abroad compared to working in the home country is larger,
    \item people's social network contains more people who already work and live abroad,
    \item and there is a lower barrier to move abroad.
\end{itemize}

In addition to matching qualitative patterns, quantitative pattern matching (i.e., model calibration) is also a desired property. However, obtaining the required quantitative data is often a challenge. Although some countries are often associated with higher brain drain through observations and anecdotal evidence, public statistics are not available on the number of emigrants from a country by education/skill levels; to the best of our knowledge. Therefore, the following methodology is employed to obtain a real-world observation regarding the number of emigrants. Since our study models the emigration of skilled people and not any people, we identified those universities in Turkey that are among the best 1000 universities worldwide according to Times Higher Education World University Rankings 2018 (available online at http://timeshighereducation.com). Alumni of the selected 16 universities are presumed to reflect the general characteristics of skilled people. The alumni information available on LinkedIn (http://linkedin.com) is utilised to calculate the ratio of alumni who live in Turkey to alumni who live elsewhere. Among the total of $622,090$ alumni, $540,841$ live in Turkey; corresponding to the ratio of $86.9\%$. This ratio is then used in the calibration of the proposed model. However, note that, the general purpose of the model is not to model Turkey in particular and data on Turkey is only used for demonstration purposes.

It is assumed that, in the base run, both $G_0$ and $G^*$ have the value of $0$ so that they serve as an intuitive reference point. Any value of $G_0$ or $G^*$ is then interpreted comparative to the state in the base run. Dunbar's number \cite{dunbar1992neocortex} dictates that a person can maintain only up to 150 stable social relations. Only a small portion of them would be in the workforce and be relevant in facilitating emigration. It is then assumed that, on average, an individual has 20 meaningful connections who might help the individual emigrate if they already work abroad. It corresponds to the value of $10$ for the parameter $linksToNodesRatio$. Furthermore, the underlying social network in the base run is assumed to be the random network model based on \cite{erdos1960}. All these suppositions leave three parameters for calibration: $threshold$, $luckCoef$, and $networkCoef$.

Total of 10,000 calibration experiments, in which the calibration parameters take up random values selected uniformly from their ranges, are performed. The lower and upper values for $luckCoef$ is set as $0.25$ and $0.75$. The lower and upper values for $networkCoef$ is set as $0$ and $2$. For $threshold$, the lower value is set as $0$ since a negative threshold is not meaningful in this model, and the upper value is set as $0.15$ since the range of other parameters cannot produce a $p$ that is greater than approximately $0.15$. These values are assigned based on the authors' views, intuition, and initial experiments; therefore, can be considered as \textit{guesstimation}.

The number of people who stay in the home country is observed for all calibration experiments. 11 runs are found to result in 869 home-stayers among the total population of 1000, corresponding to the observed real-world ratio of $86.9\%$. For these 11 runs, average values of $threshold$, $luckCoef$, and $networkCoef$ are found as approximately $0.04$, $0.45$, and $0.95$ respectively. For further validation, the average parameter values are then used to simulate 1000 repetitions. The average number of home-stayers in those repetitions is found as $872$ which corresponds to the ratio of $87.2\%$. Since the difference of $0.3\%$ is negligible, the found values are accepted as default values of $threshold$, $luckCoef$, and $networkCoef$ in the base run.

\section{Model analysis}\label{sec:analysis}

In this section, our proposed model is analysed by setting extreme parameter values, investigating tipping points, performing local and global sensitivity analyses, conducting uncertainty analysis, examining the model's robustness to different network types, and developing response surface models.

\subsection{Extreme parameter values and tipping points}

In this part, we have analysed the results of several heuristics such as setting extreme values for parameters and investigating tipping points.

\textbf{Extreme values of parameters - supportive.} Parameter values are set at their extreme values in their range so that that the environment is extremely supporting a greater brain drain. $threshold$ is set as $0$, $luckCoef$ is set as $0.75$, $G_0$ and $G^*$ are set as $0.5$, $networkCoef$ is set as $2$. $linksToNodesRatio$ is set as $20$ instead of $10$ and the $networktype$ is left as random network model. The model is simulated 10 times. Percentage of people who stayed in the home country is observed. Intuitively, we expect a relatively small portion of people to stay in the country. The mean and standard deviation is found as $24.8\%$ and $1.5\%$ respectively, confirming our intuition for the result of this extreme parameter setting.

\textbf{Extreme values of parameters - unsupportive.} Parameter values are set at their extreme values in their range so that the environment is extremely discouraging of brain drain. $threshold$ is set as $0.15$, $luckCoef$ is set as $0.25$, $G_0$ and $G^*$ are set as $-0.5$, $networkCoef$ is set as $0$. Since $networkCoef$ is set to $0$, the other parameters regarding network do not matter. The model is simulated for 50 times. Percentage of people who stayed in the home country is observed. Intuitively, we expect no people to emigrate due to the values of $threshold$, $G_0$ and $G^*$. In all runs, no people emigrated. Our intuition for the result of this extreme parameter setting is confirmed. Moreover, another set of simulations is conducted without the threshold (i.e., $threshold = 0.15$) and only $1.5\%$ of people emigrated on average. Hence, even in the absence of the barriers, the model with extremely unsupportive configuration behaves as expected.

\textbf{Tipping points.} We have performed a number of experiments to identify any tipping points in the model behaviour by trying different values for parameters. However, we were not able to obtain such tipping points where a small change in a parameter value results in much larger behaviour change in the model output. On the other hand; as more people immigrate, it helps other people to emigrate through network effects and create a compounding effect that is supportive of emigration. Although this often means an accelerating emigration, a strong tipping point is not observed.

\subsection{Local sensitivity analysis}
The purpose of local sensitivity analysis is to examine the sensitivity of the model to small changes in parameter values. In local sensitivity analysis experiments; for each parameter, the model is run for its reference value $P$ that is the value in the base run (i.e., the calibrated model), and for $P + dP$ and $P - dP$ where $d = 0.05$. As an exception, $dP$ is taken as $0.025$ for $G_0$ and $G^*$ since their reference values are $0$. Each experiment is replicated 10 times. The average number of people who stayed in the country is reported as $C$, $C^+$, and $C^-$ respectively. Table \ref{tab:lsa} presents the experiment results with corresponding relative changes in $C$ values. Sensitivity values $S^+$ and $S^-$ show the amount of change in $C$ corresponding to $0.01$ change in parameter reference value for $G_0$ and $G^*$, and show the percentage change in $C$ corresponding to $1\%$ change in parameter reference value for $threshold$, $luckCoef$, and $networkCoef$. According to the results, the proposed model is shown to be not too sensitive to the small changes in values of parameters as indicated by the small values in the last two columns of Table \ref{tab:lsa}.

\begin{table}[h]
\centering
\caption{Local sensitivity analysis}
\label{tab:lsa}\setlength\tabcolsep{4pt}
\begin{tabular}{r|r|r|r|r|r|r|}
\cline{2-7}
                                    & \textit{Ref. Value} & $C^-$ & $C$   & $C^+$ & $S^+$   & $S^-$   \\ \hline
\multicolumn{1}{|r|}{$G_0$}         & 0                   & 878.3 & 873.0 & 867.0 & -2.40   & 2.12    \\ \hline
\multicolumn{1}{|r|}{$G^*$}         & 0                   & 877.4 & 874.0 & 869.3 & -1.88   & 1.36    \\ \hline
\multicolumn{1}{|r|}{$threshold$}   & 0.04                & 861.2 & 866.2 & 874.0 & 0.18\%  & -0.12\% \\ \hline
\multicolumn{1}{|r|}{$luckCoef$}    & 0.45                & 885.5 & 876.1 & 856.0 & -0.46\% & 0.21\%  \\ \hline
\multicolumn{1}{|r|}{$networkCoef$} & 0.95                & 880.3 & 868.7 & 867.0 & -0.04\% & 0.27\%  \\ \hline
\end{tabular}
\end{table}

\subsection{Global sensitivity analysis}
In this part, we do some experiments to show the effect of different values of a parameter at their full range.

\textbf{Weight of network effects.} All parameter values except $networkCoef$ are kept same as the base run and five different values are tried for $networkCoef$: $0$, $0.5$, $1$, $1.5$, and $2$. For each value, the model is simulated for 5 times. The number of people who stay in the country is found as $976$, $947$, $870$, $778$, and $691$ on average. Therefore, it is concluded that the larger the weight of network effects, the larger the emigration rate. It also shows an accelerating speed of emigration as $networkCoef$ increases, due to the creation of a stronger emigration-supportive environment which results in an even stronger emigration-supportive environment, thus a compounding effect. 

\textbf{Overall utility difference in the base run.} In the base run, keeping all other parameters fixed, we try different values for $G_0$ and $G^*$ and observe the number of people who stay at home. Experiments are repeated five times for each $(G_0, G^*)$ combination and average results are reported in Table \ref{tab:g0gn}. Results show that as the utility difference increases against the benefit of the home country, the brain drain grows in size however is still limited by barriers.

\begin{table}[h]
\centering
\caption{Effects of $G_0$ and $G^*$ on number of emigrants}
\label{tab:g0gn}
\begin{tabular}{|r|r|r|r|r|r|}
\hline
$G_0$  \textbackslash $G^*$ & \textit{-0.5} & \textit{-0.25} & \textit{0} & \textit{0.25} & \textit{0.5} \\ \hline
\textit{-0.5}                      & 1000          & 890            & 875        & 866           & 827          \\ \hline
\textit{-0.25}                     & 914           & 903            & 889        & 851           & 831          \\ \hline
\textit{0}                         & 914           & 894            & 884        & 853           & 826          \\ \hline
\textit{0.25}                      & 907           & 897            & 880        & 843           & 815          \\ \hline
\textit{0.5}                       & 889           & 879            & 858        & 840           & 800          \\ \hline
\end{tabular}
\end{table}

\textbf{Overall utility difference in an emigration-supportive environment.} Differently from the previous one, we set $threshold$ value as $0$ instead of the default $0.04$ and $luckCoef$ as $0.75$ instead of the default $0.45$. Then, different values are set for $G_0$ and $G^*$ and the number of people staying at the home country is observed. Five experiments are conducted for each $(G_0, G^*)$ combination and average results are reported in Table \ref{tab:g0gnwtl}. In addition to confirming results from Table \ref{tab:g0gn}, it also shows that even though the home country might be advantageous in general, some agents might have higher utility abroad and the emigration rate is quite significant in the absence of barriers.

\begin{figure}[t]
\centering
\includegraphics[width=1\linewidth]{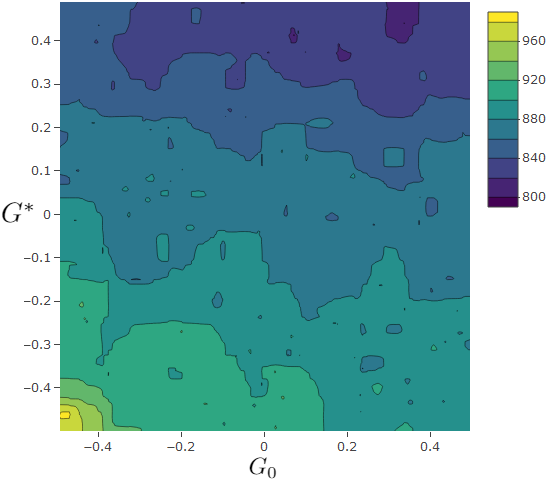}
\caption{Contour plot for effects of $G_0$ and $G^*$ on number of emigrants}
\label{fig:cont}
\end{figure}

Furthermore, 250 experiments are conducted where $G_0$ and $G^*$ take up random values uniformly from their range. The other parameters are assigned their default values in the base run. The number of people who stayed in the country is visually demonstrated in the contour plot given in Figure \ref{fig:cont}. X-axis represents $G_0$ and Y-axis represents $G^*$. When inspected, it is seen that as $G_0$ and $G^*$ increases, the number of people who stayed in the country decreases. The maximum number of people who stayed in the country is observed in the bottom-left corner where both $G_0$ and $G^*$ take their smallest values. The effect of $G^*$ on the emigration is greater than $G_0$ as shown by the approximate lines separating different colours which are approximately more horizontal rather than vertical.

\begin{table}[h]
\centering
\caption{Effects of $G_0$ and $G^*$ on number of emigrants in an emigration-supportive environment}
\label{tab:g0gnwtl}
\begin{tabular}{|r|r|r|r|r|r|}
\hline
$G_0$  \textbackslash $G^*$ & \textit{-0.5} & \textit{-0.25} & \textit{0} & \textit{0.25} & \textit{0.5} \\ \hline
\textit{-0.5}                      & 742           & 717            & 697        & 672           & 628          \\ \hline
\textit{-0.25}                     & 723           & 705            & 683        & 666           & 607          \\ \hline
\textit{0}                         & 725           & 702            & 673        & 655           & 607          \\ \hline
\textit{0.25}                      & 716           & 700            & 670        & 632           & 606          \\ \hline
\textit{0.5}                       & 724           & 705            & 655        & 626           & 610          \\ \hline
\end{tabular}
\end{table}

\subsection{Uncertainty analysis}

The purpose of uncertainty analysis is to investigate how uncertainty in parameter values and sensitivity of the model together lead to uncertainty in model outputs. Based on the previously performed local sensitivity analysis, the most sensitive parameters are found as $networkCoef$ and $luckCoef$ as indicated by relatively larger values in the last two columns of Table \ref{tab:lsa}. The values of these parameters are also naturally uncertain since they represent abstract phenomena. Therefore, uncertainty analyses are performed for them.

\begin{figure}[b]
\centering
\includegraphics[width=1\linewidth]{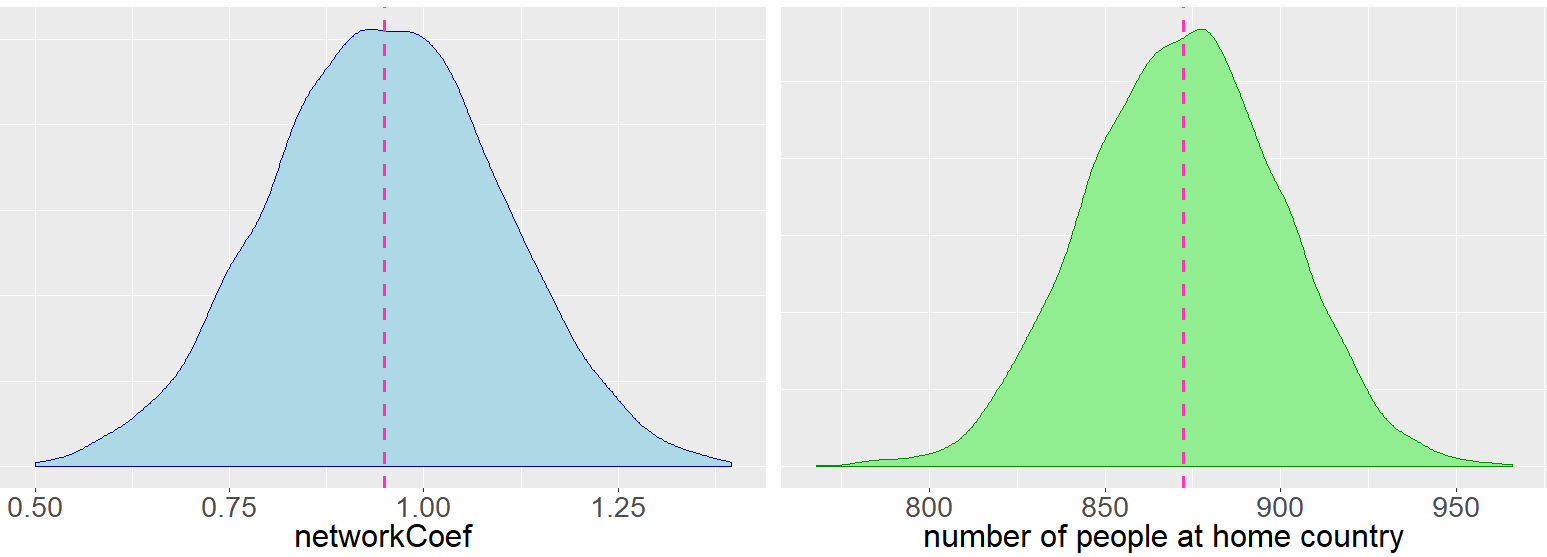}
\caption{Uncertainty analysis for $networkCoef$}
\label{fig:unc1}
\end{figure}

\begin{figure}[b]
\centering
\includegraphics[width=1\linewidth]{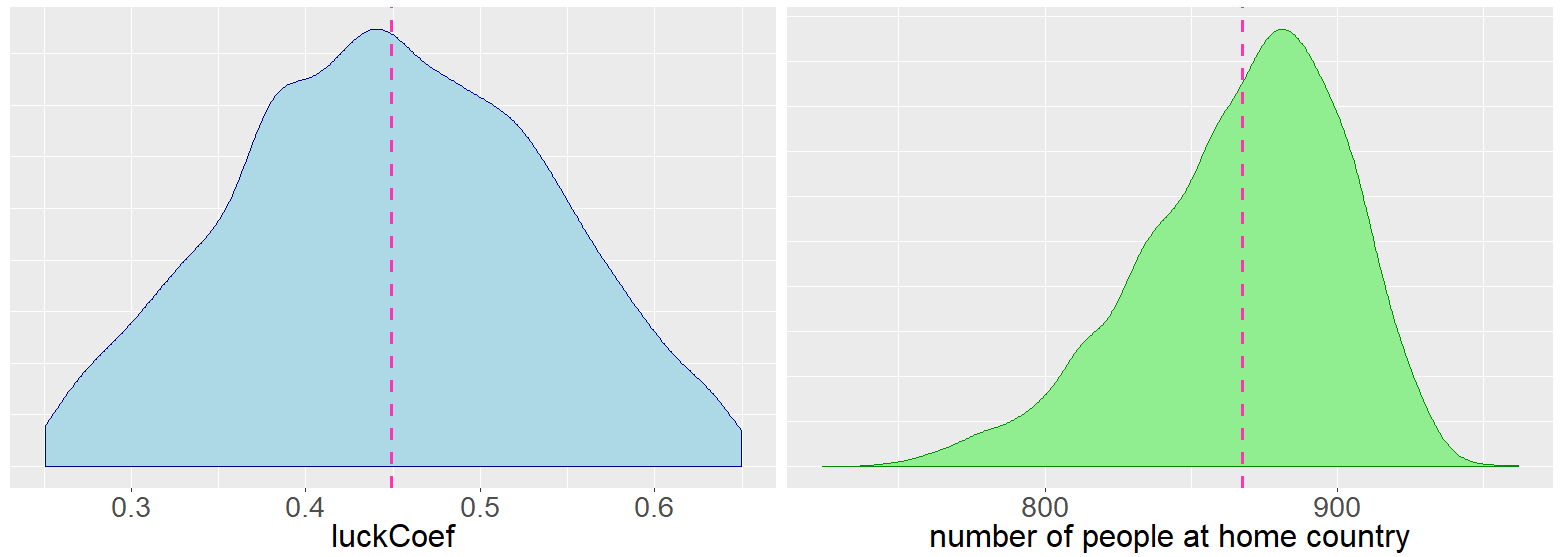}
\caption{Uncertainty analysis for $luckCoef$}
\label{fig:unc2}
\end{figure}

For total of 5000 experiments, $networkCoef$ is assigned values following a normal distribution with $\mu = 0.95$ and $\sigma = 0.15$. Values outside the $[0.5, 1.40]$ range is removed. The distributions of $networkCoef$ and simulation output (number of people who stayed home) are given in Figure \ref{fig:unc1}. Similarly, for total of 5000 experiments, $luckCoef$ is assigned values following a normal distribution with $\mu = 0.45$ and $\sigma = 0.10$. Values outside the range of $[0.25, 0.65]$ are removed. The distributions of $luckCoef$ and simulation output (number of people who stayed home) are given in Figure \ref{fig:unc2}. The figures demonstrate that the shapes of the distribution of the respective parameters and the model outputs approximately match.

\subsection{Robustness to different network types}

In order to show that our model is robust to different network types and to investigate the effects of network structures, the model is simulated with underlying preferential attachment network model \cite{barabasi1999} in addition to the default random network model. Moreover, $linksToNodesRatio$ is also included in the robustness analysis. The experiments are repeated 100 times for each setting. Values for the average number of people who stay in the home country are reported in Table \ref{tab:rob}.  The results suggest that as the network gets denser (i.e., a larger number of links), the number of people moving abroad increases since people are expected to have more contacts who already live abroad. On the other hand, according to a t-test on the given sample, there is no statistically significant difference ($p>0.1$ both when $linksToNodesRatio$ equals $10$ or $30$) in terms of brain drain when the two underlying network models are compared. Additional simulations are run with $10000$ agents instead of the original $1000$ to test whether the number of agents is too small for underlying network models to manifest their effects. However, the results were similar and a meaningful difference was not observed. It might be due to the fact that both models do not produce any local clustering or community structure where network effects might larger. Further comprehensive experiments and analysis are necessary to explore and determine the effects of various network structures on migration patterns.

\begin{table}[h]
\centering
\caption{Robustness to different networks}
\label{tab:rob}
\begin{tabular}{|r|r|r|}
\hline
\textit{network type} \textbackslash $linksToNodesRatio$ & \textit{10} & \textit{30} \\ \hline
\textit{Random network model}              & 872       & 774       \\ \hline
\textit{Preferential attachment model}     & 867       & 771       \\ \hline
\end{tabular}
\end{table}

\subsection{Response surface analysis}

In order to understand the effect of each parameter in simulation output and also to build predictive models, a linear regression model and a random forest model \cite{Breiman20015} are trained\footnote{For the linear regression model, we used ordinary least squares method. For the random forest model, number of trees is set to 1000, maximum leaf node size is set to 5, and sampling with replacement is used. All variables are considered at each splitting decision. Final prediction for a case is the mean of predictions from all trees.}. These models can be used to estimate the number of emigrants without running lengthy simulations, thus saves some effort. To generate the dataset employed in training of the response surface models, the agent-based model is simulated for 30,000 times, each time $threshold$, $luckCoef$, $G_0$, $G^*$, and $networkCoef$ taking up random values uniformly from their respective ranges. The target variable, $ratio$, is set as the percentage of people who stayed in the home country. Randomly selected 75\% of the data is used as the training set where models are trained, and the rest is used as the test set. For each model, root mean squared error (RMSE) values are reported for the models' predictive performances on the test set. In the linear model and random forest model, RMSE values are found as $8.3\%$ and $1.5\%$ respectively. Therefore, both models, particularly the random forest model, can obtain very accurate predictions without having to resort to simulations.

Given the success of the linear model in predicting the number of emigrants, an additional linear regression model is trained on all data to understand the effect sizes and significance of parameters on the number of emigrants. In contrast to the very precise but not explainable random forest model, this simple equation allows interpretation and useful insights. 
Estimated coefficients of the regression model and their respective standard errors are given in Table \ref{tab:rsf}. All variables are found to be statistically significant at $p < 0.001$. R-squared and adjusted R-squared values are found as $0.62$, approximately. Hence, the linear model explains 62\% of the variance in the ratio of nonmigrants. 

\begin{table}[h]
\centering
\caption{Coefficients}
\label{tab:rsf}
\begin{tabular}{r|r|r|}
\cline{2-3}
                                  & \textit{Estimate} & \textit{Std. Error} \\ \hline
\multicolumn{1}{|r|}{($Intercept$)} & 1.01     & 0.002      \\ \hline
\multicolumn{1}{|r|}{$threshold$}   & 1.71     & 0.011      \\ \hline
\multicolumn{1}{|r|}{$luckCoef$}    & -0.25    & 0.003      \\ \hline
\multicolumn{1}{|r|}{$G_0$}       & -0.07    & 0.003      \\ \hline
\multicolumn{1}{|r|}{$G^*$}          & -0.10    & 0.003      \\ \hline
\multicolumn{1}{|r|}{$networkCoef$} & -0.11    & 0.001      \\ \hline
\end{tabular}
\end{table}


The findings from the equation are as follows.

\begin{itemize}
    \item Larger barriers have a significant role in slowing down emigration. Such a role of $threshold$ was previously shown when supportive and unsupportive environments are discussed throughout the text.
    \item Rather than the current socio-economic status of the country, the future status has a more significant role as indicated by the larger regression coefficient of $G^*$ ($-0.1$) compared to that of $G_0$ ($-0.07$). The same result was deducted from Figure \ref{fig:cont} where $G^*$ (i.e., Y-axis) plays a more significant role in determining the number of emigrants.
    \item The larger the role (i.e., weight) of network effects on emigration, the more the brain drain. This result confirms the earlier finding in global sensitivity analysis regarding the weight of network effects.
\end{itemize}

\section{Conclusion}\label{sec:conc}
This study provides an agent-based model for emigration dynamics of skilled workers. A few abstract variables are proposed to represent the variables employed in the related literature, capturing the main factors.  The proposed computational model is validated with qualitative and quantitative pattern matching, respectively via calibration to real data and visual inspection. Various what-if analyses including extremely supportive and unsupportive scenarios, the impact of different degrees of network effects, different socio-economic trends of the countries are conducted and findings are presented. Moreover, response surface models are built to both predict and explain the brain drain dynamics. The estimated coefficients of the linear model, where all parameters are statistically significant, confirm the earlier analyses on the effects of each parameter. The main insights derived from the model analysis and directions for future research are discussed in this closing section.

Predictive linear and random forest models are developed to provide a quick way of prediction on the number of emigrants given the values of model parameters. Although the linear model performs sufficiently well, the random forest model is shown to perform better. The linear model also helps to understand the effect sizes of parameters on the number of emigrants.

The visual analysis on the effects of socio-economic trend and the estimated linear model agree that the future socioeconomic state of the country is more important than the current state in retaining the skilled workers at home.  Experiments and the linear model also indicate that lack of barriers results in large emigration even if the home country has better socioeconomic status in general, due to the fact that individual utilities might differ from the overall utility difference. 

Moreover, social network effects have a compounding effect on emigration rate since the more the people emigrate, the larger the network effects become. However, no significant difference was found when the underlying network structure is random or based on preferential attachment. On the other hand, as the network gets denser (i.e., when people have more friends on average), the emigration of skilled workers accelerates.

It is often argued that simpler models are better. In accordance with this premise, we kept our model as simple and as abstract as possible. As a consequence, many details are omitted, limitations exist, and some aspects are left for future researchers. For instance, immigration into the country (i.e., brain gain) as the return of the previous emigrants or as new immigrants are not studied. Also, population growth in the population via more births than deaths is not considered. Consecutively, the population of the source country is always non-increasing.

Future researchers may work towards transforming abstract models such as ours to applied models by replacing the abstract variables with relevant observable parameters such as salary levels, cost of living, happiness index, freedom index, and so on; and by validating their models through qualitative and quantitative methods using extensive real-world data. Such models then serve as a useful decision support tool for designing actual real-world policies. In addition, more realistic networks with high clustering coefficient and community structure can be utilized instead of the simpler models we employed. The network structure can even be estimated if network data can be collected.

Apart from operationalizing the proposed model by replacing the abstract or simplistic aspects with concrete or more complex counterparts, a process which often requires real-world data, the model can be extended in other ways. For instance, side effects for introducing too high $threshold$ values (e.g., obstructing the emigration via harsh preventive policies) can be embedded into the model where it negatively affects the utilities of agents (e.g., they become unhappy, reducing their $y$). In addition, the socioeconomic status of the country, which is modelled as a hyperparameter in this study, can be modelled to partially arise from some aggregation of the utility of individual agents at home. Addition of such feedback loops or extensions might result in the emergence of additional interesting model behaviours.

\bibliographystyle{IEEEtran}
\bibliography{IEEEabrv,ref}




%







\end{document}